\documentclass[twocolumn,superscriptaddress,secnumarabic,aps,pra,nobibnotes,groupedaddress,numerical]{revtex4-1}

\usepackage{lipsum,capt-of,graphicx}
\usepackage[squaren]{SIunits}
\usepackage{verbatim}
\usepackage{mathrsfs}
\usepackage{comment}
\usepackage{braket}
\usepackage{bbold}
\usepackage{subfigure}
\usepackage{dcolumn}
\usepackage{amsmath}
\usepackage{hyperref}
\usepackage{enumerate}
\usepackage[11pt]{moresize}
\usepackage[normalem]{ulem}
\usepackage{xcolor}
\usepackage{dsfont}
\usepackage{color}
\usepackage{times}
\usepackage{amsmath}
\usepackage{amssymb}
\usepackage{booktabs}

\newcommand{\tr} {\text{tr}}
\usepackage{lipsum}

\renewcommand{\thesubfigure}{\thefigure.\arabic{subfigure}} \makeatletter
\renewcommand{\p@subfigure}{}
\renewcommand{\@thesubfigure}{\thesubfigure.\,\,\hskip\subfiglabelskip} \makeatother

\begin{document}

\title{Quantum stabilization of photonic spatial correlations}

\author{Matteo Biondi} 
\author{Saskia Lienhard} 
\author{Gianni Blatter}
\author{Sebastian Schmidt}
\affiliation{Institute for Theoretical Physics, ETH Zurich, 8093 Z\"urich, Switzerland}

	\begin{abstract}
The driven, dissipative Bose-Hubbard model (BHM) provides a generic description of collective phases of interacting photons in cavity arrays. In the limit
of strong optical nonlinearities (hard-core limit), the BHM maps on the dissipative, transverse-field XY model (XYM). The steady-state 
of the XYM can be analyzed using mean-field theory, which reveals a plethora of interesting dynamical phenomena. For example, strong hopping combined with a blue-detuned drive, leads to an instability of the homogeneous  steady-state with respect to antiferromagnetic fluctuations. In this paper, we address the question whether such an antiferromagnetic instability survives in the presence of quantum correlations beyond the mean-field approximation. For that purpose, we employ a self-consistent $1/z$ expansion for the density matrix, where $z$ is the lattice coordination number, i.e., the number of nearest neighbours for each site.  We show that quantum fluctuations stabilize a new homogeneous steady-state with antiferromagnetic correlations in agreement with exact numerical simulations for finite lattices. The latter manifests itself as short-ranged oscillations of the first and second-order spatial coherence functions of the photons emitted by the array.
	\end{abstract}
	\maketitle
	
\section{Introduction}
Photonic systems provide an ideal platform for the study of many-body physics far from equilibrium. In particular, coupled nonlinear cavities 
allow to engineer strongly correlated and exotic states of light with interesting spatial structure. For example, non-trivial spatial order of interacting photons associated with a breaking of the translational lattice symmetry was predicted due to fermionization \cite{chang2008}, modulated pumping \cite{hartmann2010}, long-range interactions \cite{otterbach2013,jin2013}, and geometric frustration \cite{biondi2015}. While early works in this context were mostly theoretical, the recent progress in various cavity QED technologies, e.g., based on cold atoms, exciton-polaritons, and superconducting circuits, led to first experimental realizations of small photonic quantum simulators \cite{hafezi2013,Raftery2014,Eichler2015,Baboux2015,Anderson2016,Fitzpatrick2017,Fink2017} 
(for two recent reviews, see Refs.~\cite{Noh2016,Hartmann2016}). 

The experimental advances in assembling nonlinear photonic cavity systems is also a strong motivation for developing novel mathematical tools and methodology.
The key object is typically a master equation, which describes the dynamical evolution of the system density matrix $\rho$. 
Solving the master equation exactly is a formidable numerical task \cite{breuer2002}. 
In the limit of weak optical nonlinearities, it is typically sufficient to use semiclassical methods and include quantum fluctuations by accounting for Gaussian fluctuations
or by using stochastic wave function methods \cite{schleich2001}. Recently, it was proposed that the presence of a lattice may lead to emergent equilibrium behaviour even in strongly driven, dissipative systems, which then facilitates the use of standard renormalization group techniques \cite{sieberer2013,FossFeig2017}.
However, in the opposite limit of strong optical nonlinearities and weak hopping, typically neither semiclassical nor quasi-equilibrium methods provide a suitable starting point for a theoretical analysis. Under weak driving conditions, exact diagonalization and quantum-trajectories \cite{Dalibard1992,carmichael1992,Plenio1998,Daley2014} 
allow to successfully address this problem for small system sizes. Large-scale numerical methods based on tensor networks \cite{zwolak2004,Schollwoek2011,Cui2015,dorda2015,Mascarenhas2015,biondi2015}
can be applied to infinite lattices, but are mostly limited to one dimension (1D). A recently developed corner-space renormalization technique \cite{Finazzi2015} may provide an alternative also in two dimensions (2D). 

Decoupling mean-field theory allows to describe exactly local quantum fluctuations beyond semiclassical methods 
and is correct in infinite lattice dimensions, thus providing a simple tool to gain first insights into the qualitative physics at hand \cite{nissen2012,Lee2012,Ates2012,jin2013,boite2013,schiro2016,Biondi2016,Wilson2016,FossFeig2017,Biella2017}. 
Recent efforts to improve on the mean-field approximation include perturbative \cite{delValle2013,Li2014}, projective \cite{DegenfeldSchonburg2014}, cluster \cite{jin2016}, variational \cite{Weimer2015} and equations-of-motion approaches \cite{Casteels2016}. 
In Ref.~\cite{Biondi2017}, we developed a systematic expansion around the decoupling mean-field solution in powers of the inverse dimensionality parameter $1/z$ (with $z$ being the number of nearest neighbours). Such an expansion was originally developed in order to calculate ground and excited states of lattice systems in equilibrium \cite{Metzner1991,Ohliger2013,schmidt2009}. First non-equilibrium versions were discussed in \cite{Navez2010,Queisser2012,Weimer2015*2}.
In Ref.~\cite{Biondi2017}, we expanded on previous efforts by developing a self-consistent scheme to solve for the density matrix up to second order in $1/z$.
We showed, that the self-consistency condition substantially improves the results of a bare second-order expansion and compares well with large-scale numerical methods.

Here, we study the dissipative, transverse-field XY model, which describes coupled-cavity arrays in the limit of large optical nonlinearity, in order to address the role of quantum fluctuations beyond mean-field theory. A decoupling mean-field theory combined with linear stability analysis predicts a symmetry-breaking instability towards antiferromagnetic order beyond a critical value of the hopping strength $J$ \cite{Wilson2016}. In this parameter regime, no stable, homogenous steady-state exists. 
Such an antiferromagnetic instability is particularly interesting since the spin-spin couplings in the effective Hamiltonian are purely ferromagnetic. 
Using exact quantum trajectory simulations of finite lattices, one finds instead a homogeneous steady-state with antiferromagnetic spatial correlations,
rather than antiferromagnetic order of the steady-state itself. 
Unfortunately, the lattice sizes were too small ($\sim 12$ sites) to conclude about critical behaviour in the infinite system as signalled by a closing of the Liouvillian gap.

In this paper, we employ a self-consistent $1/z$ expansion to discuss 
the role of spatial correlations and quantum fluctuations in the infinite system beyond the mean-field approximation.
We show that quantum fluctuations destroy the symmetry-breaking instability and stabilize a homogeneous steady-state with antiferromagnetic correlations in agreement with exact numerical simulations. The latter manifest as short-range oscillations of the photonic spatial coherence functions, which can be measured by detecting the photons emitted by the cavity array. We also provide simple arguments to describe the origin of the antiferromagnetic correlations in the lattice. Our results demonstrate that the self-consistent version of the $1/z$ expansion provides a valid and comparably simple tool for the study of quantum fluctuations and spatial correlations in driven, dissipative systems.

In the following, we introduce the model for the driven, dissipative cavity array in Section \ref{sec:model} and describe the $1/z$ expansion in Section \ref{sec:method}. In Section \ref{sec:meanfield}, we review the results of mean-field theory and a linear stability analysis before we describe our results beyond the mean-field approximation in Section \ref{sec:qfluct}. We conclude with a brief summary and outlook in Section \ref{sec:summary}.

\section{Model}
\label{sec:model}
Our starting point is the Bose-Hubbard model
\begin{equation}
\begin{split}
\label{h_BHM}
   H & = \sum_i h_i + \frac{1}{z} 
   \sum_{ \langle ij \rangle } J_{ij} a^\dagger_i a_j, \\   
      h_i & = - \Delta\,n_i + Un_i(n_i - 1)/2 
   + f a_i + f^* a^\dagger_i
   \end{split}
\end{equation}
describing photons hopping on a lattice of nonlinear cavities, where
each cavity is described by the local Hamiltonian $h_i$ expressed in terms of
the bosonic operator $a_i$ and the associated density operator $n_i =
a^\dagger_i a_i$. Here, each site $i$ is coherently pumped with strength $f= |f|\,\exp(i\varphi)$, where $\varphi$ is the
phase of the external drive, as described by the
last term in $h_i$. In a frame rotating with the drive frequency $\omega_d$, the
cavity frequency is renormalized to $\Delta = \omega_d - \omega_c$, while $U$
is the local Kerr nonlinearity. The second
term in $H$ describes the hopping to $z$ nearest-neighbor cavities with
amplitude $J_{ij} = -J$; the additional factor $1/z$ in Eq.~\eqref{h_BHM}
ensures that the bandwidth of the photon dispersion is $2J$, independent of $z$,
and guarantees a regular limit $z\to \infty$. The dissipative dynamics for the
density matrix $\rho$ is accounted for via Lindblad's master equation,
\begin{eqnarray}
\dot{\rho} = -i[H,\rho] + \frac{\kappa}{2} \sum_i D[a_i] \rho,
\label{lindblad_full}
\end{eqnarray}
where $D[a] \rho = 2 a\rho a^\dagger - a^\dagger a\rho - \rho a^\dagger a$ and
$\kappa$ is the photon decay rate. This model can be realized in quantum
engineered settings using state-of-the-art semiconductor- \cite{carusotto2013} 
as well as superconductor technologies \cite{houck2012,schmidt2013*2,Leib2014}. 

In the limit of large on-site nonlinearity (hard-core limit $U\rightarrow \infty$), the
double occupation of lattice sites is suppressed and the local Hilbert space
cutoff $n_p$ (i.e., the maximal number of photons per site) can be restricted
to unity ($n_p = 1$). In this regime, photon operators are mapped to spin Pauli
operators $a_i \rightarrow \sigma_i^-$, $n_i \rightarrow (\sigma_i^z + \mathds{1})/2$
with corresponding ground $\ket{g_i} = \ket{0_i}$
and excited $\ket{e_i} = \ket{1_i}$ states, where $\ket{0_i} (\ket{1_i})$ denote photon Fock states with zero (one) photons 
at site $i$. In this limit, the BHM can be written as 
\begin{equation}
\begin{split}
\label{h_XYM}
   H & \!=\! \sum_i h_i + \frac{1}{z} 
   \!\sum_{\langle ij \rangle}\! J_{ij} (\sigma^x_i \sigma^x_j \!+\! \sigma^y_i \sigma^y_j \!-\! i\sigma^x_i \sigma^y_j \!+\! i\sigma^y_i \sigma^x_j),\\
   h_i & \!=\! -\Delta\, (\sigma_i^z + \mathds{1})/2 
+ |f|\cos(\varphi)\,\sigma_i^x  + |f|\sin(\varphi)\sigma_i^y.
\vspace{-2pt}
\end{split}
\end{equation}
The model Hamiltonian~\eqref{h_XYM} above resembles a spin-$1/2$ XY model (XYM) with ferromagnetic couplings between nearest-neighboring spins $J_{ij} = -J$, transverse field $\Delta$ and in-plane field $f$. Dissipation is taken into account as in \eqref{lindblad_full} with the collapse operator replacement $a_i \rightarrow \sigma_i^-$. In the rest of the paper, we will investigate the effective spin
model (\ref{h_XYM}) rather than the full BHM.

\section{$1/z$ Expansion}
\label{sec:method}
In order to solve for the nonequilibrium steady-state (NESS) of Eq.~\eqref{lindblad_full}, $\dot{\rho} = 0$, we make use of an expansion
in the inverse coordination number of the array ($\sim 1/z$). This method was introduced 
to study the equilibrium properties of Hubbard-like models \cite{Metzner1991,schmidt2009,Ohliger2013} 
and was recently extended to nonequilibrium \cite{Navez2010} and driven-dissipative systems \cite{Weimer2015*2,Biondi2017}. 
We define the reduced density matrices of one and two lattice sites 
$\rho_i=\tr_{\neq i}[\rho]$ and $\rho_{ij}=\tr_{\neq ij}[\rho]$. 
The two-sites density matrix can be decomposed into a factorizable and a correlated term as  $\rho_{ij}=\rho_{i}\rho_{j} + \rho_{ij}^c$. 
Starting from Eq.~\eqref{lindblad_full}, one
finds the dynamics of the reduced density matrix $\rho_i$,
\begin{equation}
\label{eq:rho_i}
i\dot{\rho}_i = \mathcal{L}_i \rho_i + \frac{1}{z}\sum_{j\neq i} 
\tr_{j}[\mathcal{L}^{\rm \scriptscriptstyle S}_{ij} (\rho_{i} \rho_{j} +  \rho^c_{ij})].
\end{equation}
Above, we introduced the notation $\mathcal{L}^{\rm \scriptscriptstyle S}_{ij} = \mathcal{L}_{ij}
+ \mathcal{L}_{ji}$, $\mathcal{L}_{ij} \rho = J_{ij}[\sigma^+_i \sigma^-_j, \rho]$
and $\mathcal{L}_i\rho = [h_i,\rho] + i(\kappa/2)D[\sigma^-_i]\rho$.
The equation for $\rho_i$ is coupled to the correlated part $\rho^c_{ij}$ of the two-site density matrix
that is of order $1/z$. The latter's dynamics is coupled to the three-site
term $\rho^c_{ijk}$ which is of order $1/z^2$ and so on, resulting in a systematic expansion in powers of $1/z$ \cite{Navez2010}. Such a scaling
of correlations is known from the Bogoliubov-Born-Green-Kirkwood-Yvon (BBGKY) hierarchy of statistical mechanics \cite{Weimer2015*2,Cercignani1997}. 
Including correlations $\rho^c_{ij},\rho^c_{ijk}$ and solving iteratively allows for a systematic improvement of the mean-field solution. 
This self-consistent approach is described in the Appendix~\ref{appendix:hierarchy} and it has been shown to yield accurate results when compared to exact methods \cite{Biondi2017}.

\begin{figure}[t]
\centering
\includegraphics[width=0.5\textwidth]{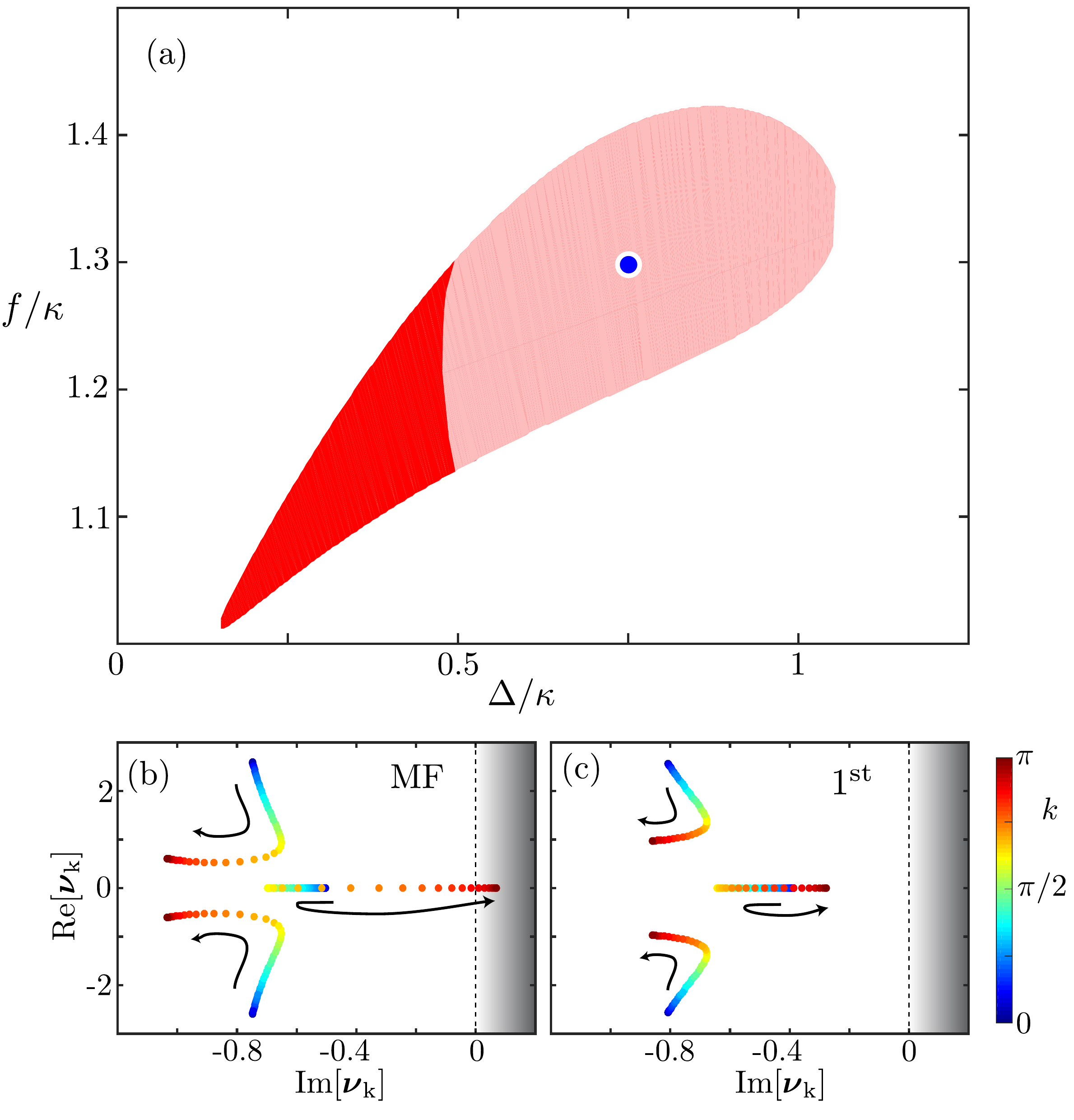}
\caption{(color online). (a) Region of mean-field instability (dark and light red) for the pumped and dissipative XYM for fixed hopping $J/\kappa= 2.25$. 
Including quantum fluctuations to first order corrections in $1/z$ stabilizes a homogenous steady-state 
with antiferromagnetic correlations for $\Delta/\kappa \gtrsim 0.5$ (light red). 
Panel (b) shows the eigenvalues of the Jacobian matrix $\boldsymbol{\nu}_k$ in the 
complex plane for $f/\kappa=1.3$ and $\Delta/\kappa=0.75$ (blue dot in (a)), 
with the $k$ values in the first Brillouin zone in color code. 
Note that at fixed $k$ the Jacobian \eqref{eq:lstab} has three eigenvalues 
(the arrows illustrate the flow in the complex plane of each eigenvalue as a function of $k$). 
The eigenvalue with the largest positive imaginary part is found for $k=\pi$ (red)  
corresponding to an antiferromagnetic instability of the homogenous mean-field steady-state. 
Panel (c) shows the eigenvalues in the complex plane for the same 
parameters as in (b), but calculated with a density matrix including $1/z$ corrections. All eigenvalues have negative imaginary part corresponding to a stable steady-state. \label{af_instab}}
\end{figure}
\section{Mean-field theory and stability analysis}
\label{sec:meanfield}
In order to solve Eq.~\eqref{eq:rho_i}, we first employ the mean-field approximation, where the connected two-site density matrix $\rho_{ij}^c$ is set to zero.
We also express the local density matrix as a vector 
$\boldsymbol{\rho}_i = [\rho_{0_i0_i},\rho_{0_i1_i},\rho_{1_i0_i}]^T$. For the steady-state with $\dot{\rho}_i = 0$, we obtain
three coupled nonlinear equations, which can be cast into the form of a single cubic equation for the density $n = (\langle\sigma_i^z\rangle + 1)/2$ 
and solved analytically. For hopping (spin-spin coupling) values $J$ larger than a critical value $J_c(\Delta,f)$ the cubic equation exhibits three real solutions where only two are dynamically stable and correspond to a low density phase $n\approx0$ and a high density (saturated) phase $n\approx1/2$. In this work, we focus on a regime of parameters with $J<J_c$ where there is only one solution to the cubic equation with an intermediate density $n$. Yet, the nonlinearity induced by the hopping $J$ gives rise to an interesting phase diagram characterized by inhomogeneous dynamical instabilities which can only be captured by studying the stability of the homogeneous density phase to perturbations with a finite momentum $\mathbf{k}$. The phase diagram of the model \eqref{h_XYM} is very rich and can also exhibit limit cycle phases, which we do not investigate here \cite{Wilson2016}. In order to address the stability of the homogeneous density phase, we perform a linear stability analysis of the mean-field steady-state $\boldsymbol{\rho}^{ss}$ and write the density matrix as $\boldsymbol{\rho}_i = \boldsymbol{\rho}^{ss} + \delta\boldsymbol{\rho}_i$,
where $\delta\boldsymbol{\rho}_i$ denotes small fluctuations on top of the mean-field. Expanding Eq.~\eqref{eq:rho_i} to linear order in the
fluctuations, one finds for each $\mathbf{k}$ in the first Brillouin zone a linear equation of the form
\begin{eqnarray}
i\frac{d}{dt} \boldsymbol{\delta\rho}_\mathbf{k} = \mathcal{J}_\mathbf{k} \boldsymbol{\delta\rho}_\mathbf{k} 
\label{eq:lstab}
\end{eqnarray}
with $\boldsymbol{\delta\rho}_\mathbf{k} = \frac{1}{\sqrt{N}} \sum_i e^{{-i \mathbf{k}\cdot\mathbf{r}_i}}\,\boldsymbol{\delta\rho}_i$. Here, $\mathbf{r}_i$ denotes the position of site $i$, $N$ is the number of sites and $ \mathcal{J}_\mathbf{k}$ the Jacobian matrix derived from the linearization,
\begin{eqnarray}
\mathcal{J}_\mathbf{k} \!=\! \left[\!\!\!\begin{array}{ccc} 
-i\kappa & - J_\mathbf{k}\phi - \tilde{\phi} & J_\mathbf{k}\phi^* + \tilde{\phi}^*\\ 
-2\tilde{\phi}^* & \Delta -i\kappa/2 - J_\mathbf{k} X & 0 \\ 
2\tilde{\phi}  & 0 & -\Delta -i\kappa/2 + J_\mathbf{k} X
\end{array}\!\!\right]
\label{eq:jacobian}
\end{eqnarray}
with $\phi = \rho_{10}^{ss}$, $X = \rho^{ss}_{11} -  \rho^{ss}_{00}$ and $\tilde{\phi} = f - J\phi$. The lattice geometry enters the Jacobian matrix through the dispersion relation $J_\mathbf{k} = \frac{1}{zN} \sum_{ij} J_{ij} e^{i\mathbf{k} (\mathbf{r}_i - \mathbf{r}_j)}$. In the 
following, we focus on the one-dimensional (1D) case with $J_\mathbf{k} = - J\cos k $, $k = 2\pi p/N$, $p=0,\dots,N-1$. The steady-state solution $\rho^{ss} $ is stable if all the eigenvalues $\boldsymbol{\nu}_k$ of the Jacobian have a negative imaginary part (at fixed $k$ the Jacobian \eqref{eq:lstab} has three eigenvalues). If one of the eigenvalues develops a positive imaginary part, the steady-state becomes unstable.

In Ref.~\cite{Wilson2016}, it was discussed that different instabilities
---which can be classified using the eigenvalue with the largest (positive) imaginary part of the Jacobian matrix---
of the steady-state solution $\rho^{ss}$ arise for positive detuning $\Delta>0$ and finite hopping $J$, with $J$ lower than the critical threshold $J_c$ for bistability. We find that for $J/\kappa \gtrsim 2$ a region characterized by a unique instability shows up, see Fig.~\ref{af_instab}(a). The red area in the Figure marks the region of the instability at fixed $J/\kappa = 2.25$ in a plane defined by pump strength $f$ and detuning $\Delta$. In the unstable regime, the eigenvalue at $k=\pi$ has the largest (positive) imaginary part (see Fig.~\ref{af_instab}(b)). The fluctuations on top of the steady-state are thus mostly antiferromagnetically ordered.

\section{Quantum fluctuations beyond mean-field}
\label{sec:qfluct}
\begin{figure}[t]
\centering
\includegraphics[width=0.35\textwidth]{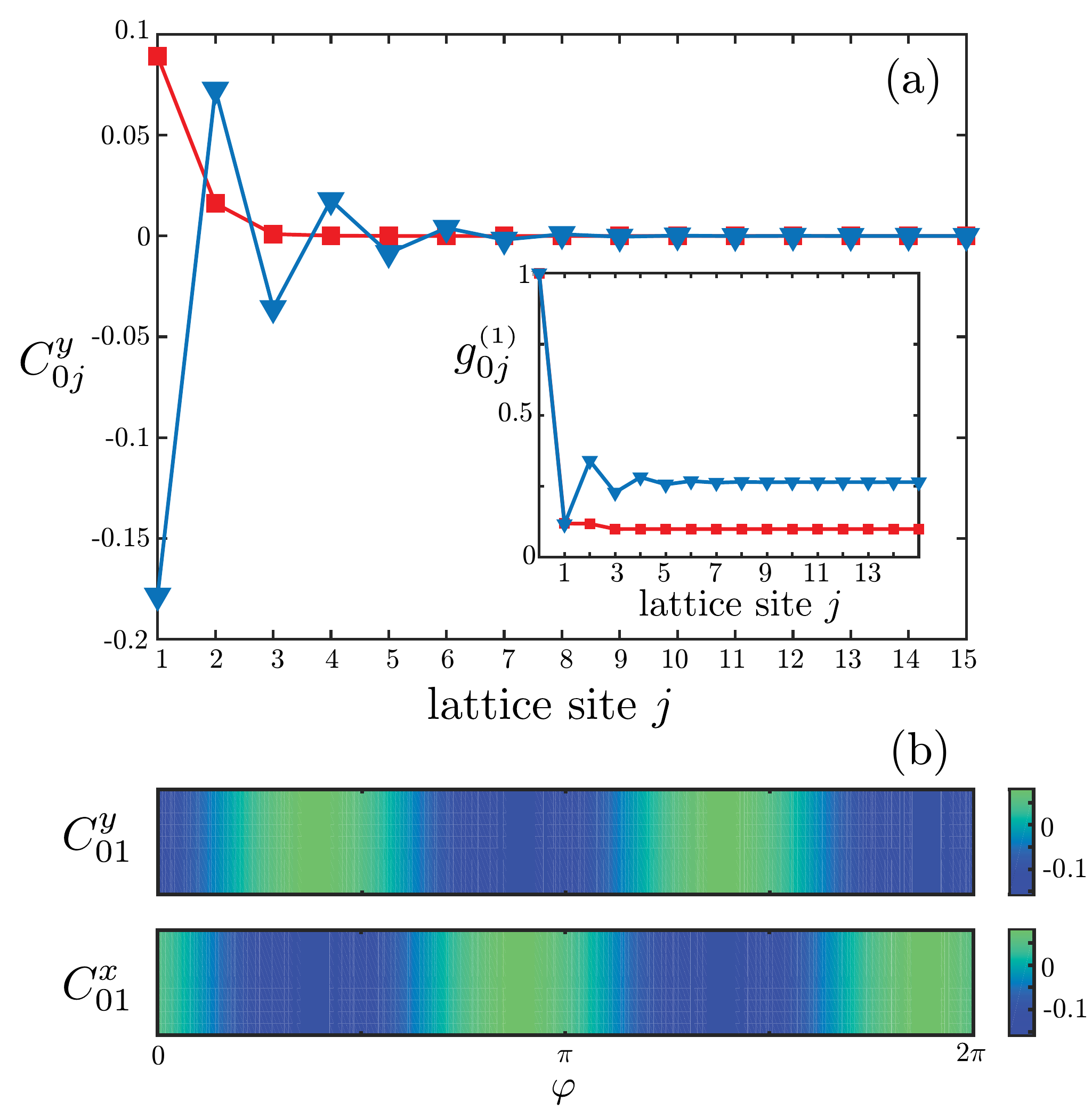}
\caption{(color online). Spin-spin correlator along the $y$ direction $C^y_{0j}$ (a) as a function of lattice 
site $j$ calculated with the $1/z$ expansion to first order at fixed drive strength $f/\kappa = 1.3$ and hopping $J/\kappa= 2.25$. 
The antiferromagnetic correlations in (a) extend further out in the lattice 
for positive detuning $\Delta/\kappa = 0.75$ (blue triangles) but not for negative detuning $\Delta/\kappa = - 0.75$ (red squares). 
The inset in (a) shows the first-order $g^{\scriptscriptstyle (1)}_{0j}$ coherence of the photons emitted by the cavities, which can be expressed in terms of the spin-spin correlation functions, see Eq.~\eqref{eq:g1}. The correlations shown in (a) are thus directly observable via photon emission spectroscopy and can be interpreted as a modulation of the homodyne signal. (b) The direction of the antiferromagnetic correlations in the $x-y$ plane can be tuned by modulating the phase $\varphi$ of the drive parameter $f= |f|\,\exp(i\varphi)$. The panel shows the nearest-neighbor correlator $C^{\alpha}_{01}$ for $\alpha=x,y$. When $C^{\alpha}_{01}<0$ the antiferromagnetic correlations also extend to larger distances as shown in panel (a).\label{fig:corr_space}}
\end{figure}

We now go beyond the mean-field description and study site-site correlations to first order in $1/z$ using the self-consistent scheme developed in \cite{Biondi2017}.
The main steps of the method are briefly outlined in Appendix \ref{appendix:hierarchy}. We find that quantum fluctuations beyond the mean-field approximation
stabilize a homogeneous steady-state in a large range of parameters, see light red area in Fig.~\ref{af_instab}(a) 
for $\Delta/\kappa \gtrsim 0.5$. In Fig.~\ref{af_instab}(c), we plot
the eigenvalues of the steady-state and find no instability. 
Interestingly, the antiferromagnetic instability of the mean-field solution is replaced by antiferromagnetic correlations of the 
new steady-state characterized by the correlation functions
\begin{eqnarray}
\label{eq:sigma_alpha}
C^{\alpha}_{0j} = \langle \sigma^{\alpha}_j \sigma^\alpha_0\rangle - \langle \sigma^\alpha_0\rangle^2
\end{eqnarray}
with $j=1,\dots,N$ and $\alpha = x,y$. 
In Figs.~\ref{fig:corr_space}(a) and (b) we show the spatial dependence of the correlators in \eqref{eq:sigma_alpha}. We find short-range antiferromagnetic correlations along the $y$ direction with $C^y_{02j - 1} < 0$ for odd sites and $C^y_{02j} > 0$ for even sites for a vanishing phase of the drive $\varphi=0$. The extent of the correlations 
is estimated by fitting the absolute value of the correlator $|C^y_{0j}|$ with a decaying exponential, i.e, $|C^y_{0j}| \propto \exp(-j/\xi)$ 
yielding a correlation length $\xi \approx 2$. 

The antiferromagnetic correlations develops only in the $x-y$ plane. Interestingly, the direction of such correlations can be tuned via the phase $\varphi$ of the external drive with $f=|f|e^{i\varphi}$ in Eq.~\eqref{h_BHM}, see Fig.~\ref{fig:corr_space}(b), where we show the nearest-neighbor correlator $C^{\alpha}_{01}$ for $\alpha=x,y$. When $\varphi=0$, the homogeneous drive is in the $x$ direction and antiferromagnetic correlations develop predominantly in the $y$ direction. For $\varphi=\pi/2$, the drive is in the $y$ direction and antiferromagnetic correlations are pronounced in the $x$ direction. It is important to note that while Fig.~\ref{fig:corr_space}(b) shows only the nearest-neighbor correlator $C^{\alpha}_{01}$, the antiferromagnetic correlations in the $x$ or $y$ direction (depending on the value of the drive phase $\varphi$) extend to larger distances, as shown in Fig.~\ref{fig:corr_space}(a). We also note that the nearest-neighbor correlators $C^{\alpha}_{01}$ oscillates roughly with a period of $\pi$ between positive and negative values. If we focus on $C^{y}_{01}$ for simplicity, we note that the minima in the oscillations are shifted with respect to $\varphi=0,\pi$; this effect is due to different harmonics in $\varphi$ and will be explained below using a minimal model with two coupled spins. 

The observed symmetry between the $x$ and $y$ direction is not accidental and can be explained as follows. In the absence of a drive, $f=0$, the Hamiltonian as well as the Lindblad dissipator are invariant under the continuous $U(1)$ symmetry $U_z(\theta) = \prod_j\exp\left[-i\theta\sigma^z_j/2\right]$. When the drive is finite, $f\neq0$, this symmetry is broken explicitly. In particular, when $\theta=\pi/2$, we have
\begin{equation}\begin{split}
\sigma_j^y & = U_z(\pi/2)\, \sigma_j^x\, U^\dagger_z(\pi/2),\\
-\sigma_j^x & = U_z(\pi/2)\, \sigma_j^y\, U^\dagger_z(\pi/2),
\label{map}
\end{split}\end{equation}
which does not leave the drive part of the Hamiltonian \eqref{h_XYM} invariant. It is simple to verify that the transformation~\eqref{map} 
is equivalent to the map $|f| e^{i\varphi} \rightarrow |f| e^{i\varphi - \pi/2}$; in other words, the unitary operator $U_z(\theta)$ together with the parametric transformation $|f| e^{i\varphi} \rightarrow |f| e^{i(\theta + \varphi)}$ leaves the Hamiltonian \eqref{h_XYM} invariant. The resulting $x - y$ symmetry is responsible for the alternation observed in Fig.~\ref{fig:corr_space}(b) between $x$ and $y$ nearest-neighbor correlators with a period of $\pi/2$. It is important to stress that the spatial correlations discussed in Fig.~\ref{fig:corr_space} have a simple interpretation in terms of the original photonic operators, namely they are related to the first-order coherence of the photons emitted by the cavities. The first-order photonic coherence function is the sum of the correlators in $x$ and $y$ direction, i.e., 
\begin{eqnarray}
\label{eq:g1}
g^{\scriptscriptstyle (1)}_{0j} = \langle a^\dagger_0  a_j\rangle = \frac{1}{4}(C^{x}_{0j} + C^{y}_{0j}) + \langle \sigma^+_0\rangle \langle \sigma^-_j\rangle.
\end{eqnarray}
Consequently, antiferromagnetic correlations in either $x$ or $y$ direction manifest as spatial oscillations of the first-order coherence function (see inset in Fig.~\ref{fig:corr_space}(a)). These oscillations can be interpreted as a modulation of the homodyne signal for $C^y_{0j}$. The results discussed in this paper are thus directly observable via photon emission spectroscopy.

Another interesting effect manifest in Fig.~\ref{fig:corr_space}(a) is the different behavior of the correlator $C^{y}_{0j}$ depending on the value of detuning $\Delta$. We find that for negative detuning $\Delta<0$ the correlator in the $y$ direction exhibits ferromagnetic correlations extending further out in the lattice, i.e., the correlations in the $y$ direction change from ferromagnetic to antiferromagnetic depending on the sign of the drive detuning, see Fig.~\ref{fig:corr_space}(a). This different behavior can be qualitatively understood as follows: for negative detunings $\Delta<0$ the drive is mostly resonant with symmetric-like superposition states of the cavities, i.e., having the form $\ket{\psi_S} \sim (\ket{e_1,g_2,\dots} + \ket{g_1,e_2,\dots} + \dots )$ since their energy is lowered by the hopping $J$ with respect to the bare cavity frequency $\omega_c$. Here, $\ket{e_i},\ket{g_i}$ denote the eigenstates of $\sigma^z_i$ introduced above Eq.~\eqref{h_XYM}. It is simple to show that translating to the spin language $\ket{\psi_S} \sim (\ket{\uparrow_1,\uparrow_2,\dots} + \ket{\downarrow_1,\downarrow_2,\dots} + \dots)$, i.e., a ferromagnetically correlated state. Here, $\ket{\uparrow_i},\ket{\downarrow_i}$ denote the eigenstates of $\sigma^y_i$. Conversely, for positive detuning $\Delta>0$ the drive is mostly resonant with antisymmetric-like superposition states of the cavities, i.e., having the form $\ket{\psi_A} \sim (\ket{e_1,g_2,\dots} - \ket{g_1,e_2,\dots} + \dots )$ since their energy is increased by the hopping $J$ with respect to the bare cavity frequency $\omega_c$. One then finds $\ket{\psi_A} \sim (\ket{\uparrow_1,\downarrow_2,\dots} + \ket{\downarrow_1,\uparrow_2,\dots} + \dots)$, i.e., an antiferromagnetically correlated state. This explains why the correlator along the $y$ direction exhibits ferromagnetic (antiferromagnetic) behavior for negative (positive) detuning. We tested this argument by verifying that indeed 
the dependence of $C^y_{0j}$ on detuning reverses by changing the sign of $J$. 

\begin{figure}[t]
\centering
\includegraphics[width=0.35\textwidth]{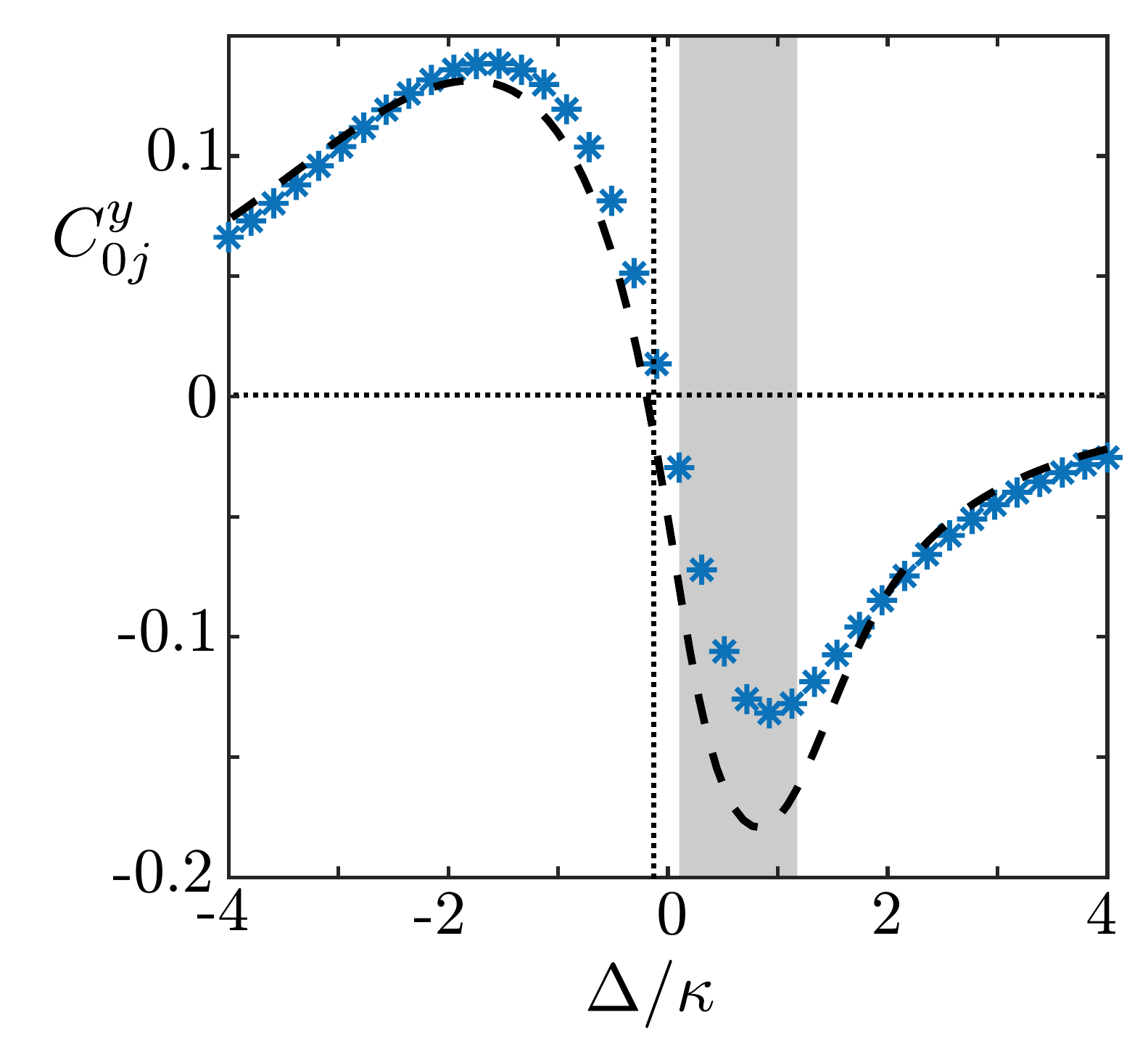}
\caption{(color online). Ferromagnetic--antiferromagnetic (F--AF) crossover in the nearest-neighbor spin-spin correlator along the $y$ direction $C^y_{01}$ 
in the steady-state as calculated with the $1/z$ method to first order (line) 
and with exact diagonalization (symbols) for a 1D lattice of $N=6$ sites. 
The correlator is shown as a function of detuning $\Delta/\kappa$ at fixed drive strength 
$f/\kappa = 1.3$ and hopping $J/\kappa=2.25$. The vertical dotted line marks the value $\Delta \approx 0$ where the correlator changes sign (see text). The grey shaded region marks the parameter regime, where the new steady-state is established (shaded area in Fig.~\ref{af_instab}).
\label{fig:F_AF_trans}}
\end{figure}

We further analyze the dependence of the nearest-neighbor correlators on the pump--cavity detuning $\Delta$ in Fig.~\ref{fig:F_AF_trans}. We compare our results with exact numerical simulations for a small finite system with only 6 sites and find good agreement. The ferromagnetic-antiferromagnetic crossover in $C^{y}_{01}$ can be understood analytically using a minimal model of just two coupled spins. Using perturbation theory in $f/\kappa$ (see Appendix~\ref{expansion_dimer}) and simplifying the resulting expressions in the large detuning limit $|\Delta|\gg \kappa, J$, we find 
\begin{equation}
C^y_{01} \approx - \frac{J|f|^2}{\Delta^3}\left[\cos(2\varphi) - (3\kappa/\Delta)\sin(2\varphi)\right].
\label{C_app}
\end{equation}
This simple result further explains the dependence of the $y$ correlator on the sign of the detuning as well as on the phase of the drive with period $\pi$ found in Fig.~\ref{fig:corr_space}(b). Note, that the contribution in Eq.~\eqref{C_app} proportional to $\sin(2\varphi)$ is of higher order in $\kappa/|\Delta|$; we included it explicitly in our result since it is responsible for the shift in the oscillations minima with respect to $\varphi=0,\pi$ observed in Fig.~\ref{fig:corr_space}(b). We note that by the symmetry argument outlined above the same analysis can be carried out for the $x$ component. Finally, we remark that an apparently similar ferromagnetic to antiferromagnetic (F-AF) crossover was also reported in Ref.~\cite{mendozaArenas2015} based on tensor network simulations for one-dimensional arrays; this result, however, concerned the spin correlator along the $z$ direction, which maps to the second-order coherence of the photons emitted by the cavities and manifests an opposite dependence on detuning $\Delta$. This effect can also be explained in terms of the selective excitation of the symmetric superposition state depending on the sign of detuning $\Delta$, see Ref.~\cite{Biondi2017}.

\section{Summary and conclusion \label{sec7}}
\label{sec:summary}
In summary, we applied a self-consistent $1/z$ expansion 
to study the nonequilibrium steady-state of the pumped and dissipative XY model beyond the mean-field approximation. 
We have shown that quantum fluctuations to order $1/z$ suppress an antiferromagnetic instability 
in part of the phase diagram and stabilize a homogeneous steady-state with antiferromagnetic correlations instead. We have provided simple arguments to describe the origin 
of the antiferromagnetic correlations in the lattice, which we confirmed by analytic calculation for a minimal model of two coupled spins. 
Our results are consistent with exact numerical methods based on tensor networks and quantum trajectories. 
While the latter require rather heavy computational efforts, our method can be carried out 
with modest computational resources. This motivates further simulations of dissipative spin chains \cite{Mascarenhas2016,rota2017},
more complex systems in (possibly) higher lattice dimensions (2D, 3D) such as Rydberg polaritons, exciton-polaritons, 
trapped ions and superconducting transmon qubits.

\section{Acknowledgements}

We acknowledge support from the Swiss National
Science Foundation and the National Centre of Competence in Research `QSIT--Quantum
Science and Technology'.

\appendix

\section{Hierarchy equations and self-consistent iteration scheme \label{appendix:hierarchy}}
Starting from Eq.~\eqref{lindblad_full}, we obtain the equation of motion for the reduced density matrices up to order $1/z$ \cite{Navez2010,Queisser2012}, i.e.,
\begin{widetext}
\begin{subequations}
\label{eq:appendix_sys_rho}
\begin{align}
\label{eq:appendix_sys_loc}
i\dot{\rho}_i = &\,\, \mathcal{L}_i \rho_i + \frac{1}{z}\sum_{j\neq i} 
\tr_{j}[\mathcal{L}^{\rm \scriptscriptstyle S}_{ij} (\rho_{i} \rho_{j} +  \underline{\rho^c_{ij}})],\\
\label{eq:appendix_sys_conn}
i\dot{\rho}_{ij}^c = &\,\, \mathcal{L}_i \rho_{ij}^c 
+ \frac{1}{z}\mathcal{L}_{ij} (\rho_{i}\rho_{j} + 
\underline{\rho^c_{ij}}) - \frac{\rho_i}{z}
\tr_i[\mathcal{L}^{\rm \scriptscriptstyle S}_{ij} (\rho_{i}\rho_j + \underline{\rho^c_{ij}})] +  
\frac{1}{z}\sum_{k\neq ij} \tr_k[\mathcal{L}^{\rm \scriptscriptstyle S}_{ik}(\underline{\rho_{ijk}^c} 
+ \rho_{ij}^c\rho_{k} + \rho_{jk}^c \rho_{i})] + (i \leftrightarrow j)%
\end{align}
\end{subequations}
\end{widetext}
with $\mathcal{L}^{\rm \scriptscriptstyle S}_{ij} = \mathcal{L}_{ij}
+ \mathcal{L}_{ji}$, $\mathcal{L}_{ij} \rho = J_{ij}[a_i^\dagger a_j, \rho]$
and $\mathcal{L}_i\rho = [h_i,\rho] + i(\kappa/2)D[a_i]\rho$, see main text. 
In the mean-field limit of infinite coordination number ($z\rightarrow\infty$) all connected density matrices are zero and 
one only needs to solve Eq.~\eqref{eq:appendix_sys_loc}, which is nonlinear and can have multiple solutions. 
However, in order to account for spatial correlations, one needs to evaluate the density matrix to higher order in $1/z$ and also
solve the equations of motion for the connected density matrices. In a first step, we make use of the scaling hierarchy $\rho_{i_1,i_2,\dots,i_s}^c = \mathcal{O}(1/z^{s-1})$
and keep on the r.h.s of each equation only terms up to order $1/z^{s-1}$, where $s$ is the number of lattice sites
in the connected density matrix on the l.h.s.~of each equation (i.e., we neglect the underlined terms). 
The resulting system of equations is then closed and 
can be solved self-consistently (for details of the self-consistency scheme, see \cite{Biondi2017}).

\section{Perturbative expansion for the dimer model \label{expansion_dimer}} 
We start from the master equation \eqref{lindblad_full} for a system described by the XYM in \eqref{h_XYM} with $N=2$ sites and the associated four basis states $\{\ket{G}\equiv \ket{gg},\ket{L}\equiv\ket{eg},\ket{R}\equiv\ket{ge},\ket{E}\equiv\ket{ee}\}$. 
We solve the resulting system of equations perturbatively by expanding the matrix elements $\rho_{GG} = \braket{G|\rho|G}$ etc. in powers of $f/\kappa$. To leading order we obtain
\begin{equation}
\begin{split}
\rho_{LG} & = \rho_{RG}= \frac{f}{\Delta_J},\\
\rho_{LL} & = \rho_{RR} =  \frac{|f|^2}{|\Delta_J|^2}, \\
\rho_{EL} & = \rho_{ER} =  \frac{f|f|^2}{\Delta_\kappa |\Delta_J|^2},\\
\rho_{EG} & =  \frac{f^2}{\Delta_\kappa \Delta_J},\\
\rho_{EE} & =  \frac{|f|^4}{|\Delta_\kappa|^2 |\Delta_J|^2}.
\end{split}
\label{sys_equations_sol}
\end{equation}
with $\Delta_\kappa=\Delta + i\kappa/2$ and $\Delta_J=\Delta_\kappa + J/2$.
The observables in \eqref{eq:sigma_alpha} are expressed in terms of these matrix elements as
\begin{equation}
C^y_{01} = \frac{1}{2}\frac{\rho_{LL} -4\rho_{EG}}{\tr[\rho]} - 4\frac{\left(\text{Im}[\rho_{LG}] + \text{Im}[\rho_{EL}\right)^2}{\tr[\rho]^2}
\label{correlators_app}
\end{equation}
with $\tr[\rho] = 1 + 2\rho_{LL} + \rho_{EE}$. Inserting the solutions \eqref{sys_equations_sol} in Eq.~\eqref{correlators_app} and expanding the resulting expression to leading order in $f/\kappa$, we obtain the analytic results for the correlators discussed in the main text.

\bibliography{BHz_refs}

\end{document}